\documentstyle[times,aps,pra,preprint,psfig]{revtex}
\begin{document}
\frenchspacing
\title{Kinetic Theoretical Foundation of Lorentzian Statistical Mechanics}
\author{Rudolf A. Treumann}
\address{Max-Planck-Institute for extraterrestrial Physics, Postfach 1603, 
D-85748 
Garching, Germany 
and \\
International Space Science Institute, Hallerstrasse 6, CH-3012 Bern, 
Switzerland \\
(tre@mpe.mpg.de and treumann@issi.unibe.ch)}
\draft
\maketitle
\begin{abstract}
A new kinetic theory Boltzmann-like collision term including correlations is 
proposed. In 
equilibrium it yields the one-particle distribution function in the form of a 
generalised-Lorentzian resembling but not being identical with the so-called 
$\kappa$ distribution frequently found in collisionless turbulent systems like 
space plasmas. 
We show that this distribution function satisfies a generalised $H$-theorem, 
yields an expression for the entropy that is concave.
Thus, the distribution is a `true' thermodynamic equilibrium distribution,  
presumably valid for turbulent systems. In 
equilibrium it is possible
to construct the fundamental thermodynamic quantities. This is done for an ideal 
gas only.
The new kinetic equation can form the basis for obtaining a set of hydrodynamic 
conservation laws and construction of a generalised transport theory for 
strongly correlated states of a system. 
\end{abstract}


\section{Introduction}
Statistical mechanics is based on kinetic theory. Classical statistical 
mechanics
derives from Boltzmann's equation which is the evolution equation for
the one-particle phase-space distribution function $f({\bf p},{\bf q},t)$.
The Boltzmann equation itself results from the BBGKY hierarchical correlation
expansion of the Liouville equation where the particular collision term
is assumed to contain short range interactions only and all orders of
correlations are neglected. In kinetic equilibrium the Boltzmann collision term
has as solution the celebrated Maxwell-Boltzmann distribution function $f_B$ as
the basic distribution function of classical equilibrium statistical
mechanics and thermodynamics. These equilibria are characterised by the
action of Boltzmann's $H$-theorem (e.g., \cite{hua87}) proving that the time 
evolution 
of the thermodynamic system tends towards maximum Boltzmannian entropy 
${\cal S}_B[f_B]$. Both $H$-theorem and entropy are functionals of the Boltzmann
distribution, ${\cal S}_B[f_B]$ being the fundamental thermodynamic function.
In the past, generalisations of this entropy to quantum statistics of
Fermions and Bosons have been very successful. In addition other quantum 
generalisations have been proposed to fractional statistics of anions
by Haldane \cite{hal91,wu93,raj95} and others. These generalisations were
mostly based on probability counting statistics. 

The function $f_B$ has been widely used without justification 
to even describe collisionless equilibria or as the initial state of the 
evolution of collisionless systems.
Final states of such systems are often described in terms of a perturbation 
analysis as solutions of a velocity space diffusion process governed by the 
Fokker-Planck equation for which sometimes it can be demonstrated that a 
$H$-theorem similar to Boltzmann's exists as well. Such a situation as in 
quasilinear theory can be taken as justification for the validity of the 
perturbation approach. However, in collisionless systems containing strong 
correlations and consequently exhibiting interactions on all scales typical for 
strongly turbulent systems and phase transitions the Fokker-Planck perturbation 
approach may become invalidated. Such systems sometimes cannot be subordinated 
to Boltzmannian statistical mechanics. Moreover, measurements of velocity space 
and energy distribution functions in realistic natural systems frequently 
exhibit properties that are distinctly different from diffusive Fokker-Planck 
solutions \cite{cri88,cri91,lin95}. The observed (classical) distribution functions 
frequently asymptotically exhibit extended high energy tails of a certain nearly 
constant slope $-\kappa$, with $\kappa$ an arbitrary positive real number that 
is limited 
from below. Such distributions have been tentatively named $\kappa$ 
distributions and have been used in the past for interpretational purposes 
\cite{vas68} as well as for formal derivations \cite{col95,rey98}, investigation 
of the dispersive properties of $\kappa$-plasmas \cite{sum91,mac95} and 
calculation of their thermal fluctuation levels \cite{mey89,mac98}. 

All these facts lead to the question if not in collisionless systems equilibrium 
states may (possibly temporarily) evolve that are distinctly different from 
Boltzmann equilibria. In fact, given a system containing some sufficiently large 
amount of free energy and being nearly collisionless, i.e. having a certain very 
small collision frequency $\nu_c$, the evolution of the system will during its 
initial state be entirely independent of the presence of collisions between its 
particle constituents. Instead it will undergo an instability, i.e., the free 
energy will drive some of the fundamental eigen modes of the system unstable. As 
a consequence, the amplitudes of these modes will grow from their thermal level 
until they assume finite values, and the system will evolve into some previously 
non-existent structure. Since collisions are very rare for times 
$t<\tau_l\sim 1/\gamma_l\ll \tau_c\sim 1/\nu_c$ comparable to 
the linear time $\tau_l\sim 1/\gamma_l$, with 
$\gamma_l$ the linear growth rate, the system behaves initially linearly. The 
eigen-mode amplitudes grow exponentially until becoming large enough that the 
linear 
approximation breaks down. The system then enters into its non-linear state. 

\begin{figure}[ht]
\centerline{\psfig{figure=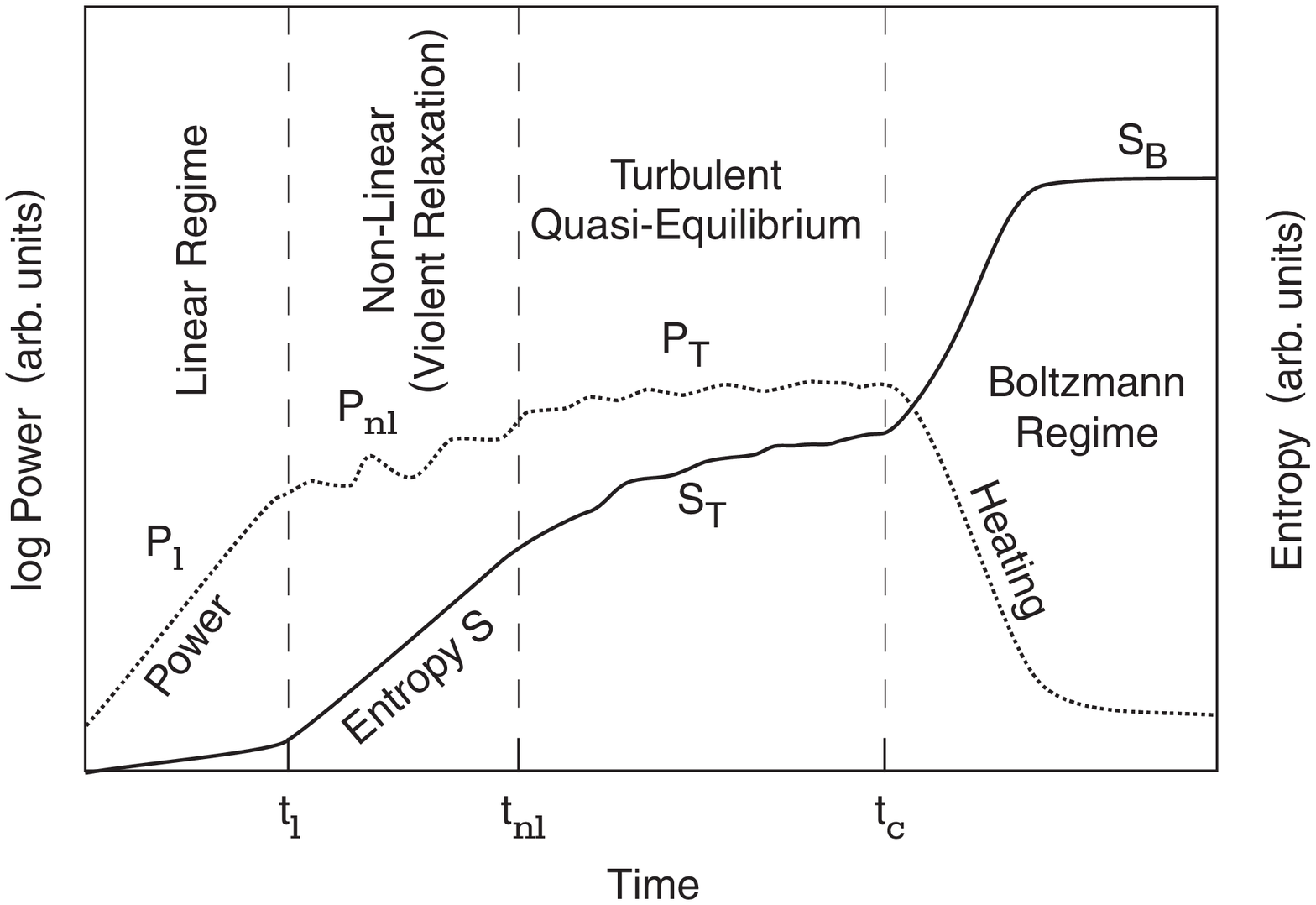,width=12.6cm,clip=}}
\caption{Evolution of the power and entropy in a nearly collisionless system 
initially containing a substantial amount of free energy. The system evolves 
through four different stages, the linearly unstable state, the nonlinearly 
unstable state, non-linear quasi-equilibrium into the final collisional thermal 
equilibrium. Pure stochasticity is reached only on this final stage when 
Boltzmann statistical mechanics holds. The non-linear quasi-equilibrium has 
slowly increasing entropy only and contains turbulent interactions on all 
scales. During this state the system may be described by non-stochastic 
processes and unconventional statistics taking into account long range 
interactions and correlations.\label{f1}}
\end{figure}

Let us assume that this state is quickly reached such that collisions are still 
unimportant. The further evolution of the system will then be determined by 
various non-linear effects like wave-wave interactions, wave-particle and 
non-linear wave-particle interactions, generation of anomalous collisions when 
the particles are scattered at the waves etc. In short, the system becomes 
turbulent.  The large-amplitude eigen-modes excite sideband waves, local 
inhomogeneity is introduced, and the system evolves into a state of coexistence 
of many different  mutually interconnected scales affecting each other. The 
total power in the system will, in this state, start to behave  
fluctuating, possibly still growing and reorganising the 
system continuously. Assume the non-linear time still being much shorter than 
the collision time, $\tau_{nl}\ll\tau_c$. For collisions still being negligible, 
this state may last for a certain possibly long turbulent time 
$t\sim\tau_T<\tau_c$. Such an evolution is sketched in Figure \ref{f1}.

The intermediate nonlinearly turbulent state represents a thermodynamic 
quasi-equilibrium state different from the purely stochastic thermodynamic 
equilibria described by the correlation-free Boltzmann statistical mechanics. 
Since the state is a quasi-equilibrium state, it should possess an equilibrium 
distribution function $f({\bf p},\kappa)\neq f_B({\bf p})$ that is different 
from $f_B$, and in addition its entropy should be approximately constant such 
that it assumes an intermediate approximate maximum for the time interval 
$\tau_{nl}<t\sim\tau_T<\tau_c$. Clearly, when the final state of the system is 
approached 
at time $t\sim\tau_c$, the system becomes collisional, and binary collisions 
will destroy the intermediate turbulent quasi-equilibrium. After $t>\tau_c$, 
stochasticity dominates over the turbulent correlations, and the power contained 
in the turbulent fluctuations is readily converted into heat. The system then 
settles at its ultimate thermal equilibrium that is described by ordinary 
Boltzmann statistical mechanics. 

The free parameter $\kappa$ introduced above heuristically describes the 
deviation of the turbulent quasi-equilibrium distribution function from the 
Boltzmann distribution. It contains the information about the actual turbulent 
processes and correlations in the system. In the absence of correlations and 
scale-invariance it will behave in a way that the quasi-equilibrium distribution 
approaches the Boltzmann distribution. The motiviation for introducing $\kappa$ 
follows the experimental literature and is thus purely traditional. However, as 
will be shown below, it turns out to be nevertheless practical. In a real 
turbulent system the entropy will still tend to increase slowly due to weak 
internal turbulent dissipation processes, but in an approximate theory we may 
assume that this increase is slow enough for the state to be considered 
stationary. These assumptions allow to ask for the possible form of the 
quasi-equilibrium distribution function $f({\bf p},\kappa)$ and for the 
statistical mechanics of the equilibrium state described by it.

The existence of a non-Boltzmannian statistical mechanics has been suspected 
since 
the observation of L\'evy flights (for an account of the current literature on 
this subject cf., e.g., \cite{shl93}). L\'evy flights have been suspected to 
affect plasma transport and diffusion. The random walks caused by L\'evy flights 
have been shown to possess an approximate distribution function resembling the 
$\kappa$ distribution \cite{tre97} though no ultimate theory of L\'evy flights 
is available yet. Beginning with Renyi \cite{bal56,ren70} intuitive but formal 
extensions of Boltzmann's theory of the statistical entropy ${\cal S}$ have been 
suggested in order to find different concepts for the entropy as the basic 
thermodynamic function of a modified statistical mechanics. A whole family of 
such formal generalisations could be imagined not all of them being of physical
relevance. More recently \cite{tsa88} a particular heuristic simplification of 
Renyi's entropy has been proposed 
which seems to deserve some physical applicability but still lacks a basic 
physical or probabilistic foundation. Among the applications proposed are 
diffusion \cite{zan95}, hydrodynamic turbulence \cite{bog96} and others. It has, moreover,
been shown that the thermodynamic formalism could be extended to this
kind of entropy \cite{bog96}, and the fluctuation-dissipation
theorem \cite{raj96} could be formulated as well. A disadvantage of
this theory is that it apparently allows for a violation of the second law
when entropic growth seems to be inhibited. Such a behaviour is
generally considered to be rather unphysical. Its advantage over Renyi's 
original proposal is its formal 
simplicity. 

In this communication we construct the one-particle kinetic equivalent of the 
Boltzmann equation which forms the physical basis for a generalised-Lorentzian
statistical mechanics. We demonstrate that in thermodynamic
equilibrium its solution is the real and positive definite 
generalised-Lorentzian phase 
space distribution 
function
\begin{equation}
f({\bf p},\kappa)=A_\kappa\left[1+\frac{\beta}{\kappa}(\epsilon_{\bf 
p}-\mu)\right]^{-
\kappa},
\label{eq:fkappa}
\end{equation}
where $\epsilon_{\bf p}=({\bf p-p}_0)^2/2m$ is the (non-relativistic) particle 
energy,
$\mu$ the chemical potential that is related to the fugacity $z=\exp\beta\mu$, 
and 
$A_\kappa$ is a certain normalisation constant (extension to the relativistic or 
to non-ideal cases including external force fields is then straightforward). The 
Lagrangean multiplier $\beta(T)$ is a function of the 
temperature $T$ of the system, which is the physical observable, $\kappa$ is a 
positive real valued but otherwise arbitrary number that can assume any positive 
value $\kappa\in {\textsf{\textbf R}}$. For $\kappa\to \infty$, the above 
distribution smoothly approaches the Boltzmann function $f_B$. 

\section{The Lorentzian Collision Term}
The final step in the BBGKY kinetic theory of solution of the exact Liouville 
equation for a many particle system is the one-particle kinetic equation for the 
one particle distribution function $f_1({\bf p}_1)$
\begin{equation}
{\cal L}f_1={\cal C}(f_1,f_2)
\label{eq:liou}
\end{equation}
where ${\cal L}\equiv \partial_t+({\bf p}_1/m)\times\nabla_1+{\dot{\bf 
p}}_1\cdot\partial_{{\bf p}_1}$ is the one-particle Liouville (Boltzmann) 
operator acting in the one-particle phase space. The collision operator ${\cal 
C}$
on the right-hand side of Equation (\ref{eq:liou}) depends on the one and 
two-particle 
distribution functions $f_1, f_2$, respectively, as well as 
on the particle number $N$. It is a very 
complicated functional that in principle traces back through the entire 
hierarchy to the original $N$-particle Liouville equation. In order to arrive at 
any physically useful solutions it must be approximated in some way. The most 
widely used approximations are Boltzmann's hypothesis of statistical 
independence of the interactions in any kind of collisions and the collisionless 
assumption of Vlasov. The latter leads to the neglect of the collision term, 
attributing all interactions to the action of the fields on the one-particle 
distribution function on the left-hand side of Eq. (\ref{eq:liou}). This 
assumption is very strong because even in the absence of any binary one-particle 
collisions, the action of the fields and very weak higher-order collisions on 
the level of 
the two-particle equation in the BBGKY hierarchy may result in a non-vanishing 
one-particle collision term ${\cal C}$. This term may contain residual 
correlations and may thus not necessarily be of the correlation-free 
statistically independent Boltzmann form proposing that ${\cal C}_B$ is a 
functional of the product of two one-particle distribution functions only:
\begin{equation}
{\cal C}_B={\cal C}_B[f_1(1)f_1(2)].
\end{equation}
The form $f_1(i)$ stands for $f_1({\bf p}_i,{\bf x}_i)$, where the index $i$ 
designates the number of the particle involved whose position in momentum and 
configuration space are ${\bf p}_i, {\bf x}_i$. Its explicit form is
\begin{equation}
{\cal C}_B=\int{\rm d}\Omega\frac{{\rm d}\sigma}{{\rm 
d}\Omega}\left\{f_1(1^\prime)f_1(2^\prime)-f_1(1)f_1(2)\right\}\equiv\int{\rm 
d}\Omega\frac{{\rm d}\sigma}{{\rm d}\Omega}\{F[1^\prime 2^\prime]
-F[12]\}.
\label{eq:collbol}
\end{equation}
The prime indicates the particles after the collision, with $\sigma$ the 
collisional cross-section, and $\Omega$ the solid angle of the collision. The 
cross-section depends on the elementary physics of the collision process and has 
to be determined separately for the individual process in question. 
In central force fields, it becomes 
the Boltzmann-Rutherford cross section. For purely field-mediated interactions 
it is a  complicate function. However, in order to determine the 
distribution function $f_B$ in thermal 
equilibrium, explicit knowledge of the differential 
cross-section is not needed. All that is required is that ${\rm 
d}\sigma(\Omega)/{\rm d}\Omega >0$ is positiv definite. This holds for any 
real physical process.

The molecular chaos assumption underlying Boltzmann's theory refers to the 
absence of all 
interactions between the particles other than binary with two-particle 
correlation function $F[12]\approx f_1({\bf p}_1)f_1({\bf p}_2)$. It
 thus reduces to $F[12]$ to
the product of the two one-particle distribution functions $f_1(1), f_1(2)$. In 
such an approach the binary interactions dominate on all scales and for all 
times, 
an assumption that has turned out very fruitful in classical statistical 
mechanics. However, as mentioned above, under 
certain conditions like scale-invariance this assumption may become violated. 
Self-consistent field 
fluctuations may lead to long-range correlations in the system and clustering of 
particles in groups that behave 
similarly on many different scales. Under such conditions molecular chaos will 
be replaced by chaos on 
larger scales that are dominated by correlations among the particles. Clearly, 
such interactions  depend 
crucially on the particular kind of correlation induced, and the general 
behaviour of the system  
becomes very difficult to investigate. 

There are two primary effects caused by 
such correlations. The first is a change in the differential cross-section. 
More important for our purposes here is the change induced in
the two-particle correlation function $F[12]$ in the collision 
integral (\ref{eq:collbol}). Without explicit knowledge about the particular 
turbulent process it seems that little more could be said. However, in a first 
approach to a statistical theory one may follow Boltzmann's approach and assume 
that even in the turbulent regime the one-particle distribution 
function  still satisfied the lowest-order kinetic equation of the BBGKY 
hierarchy, i.e., Boltzmann's equation (\ref{eq:liou}), but now with 
different (turbulent) collision term
\begin{equation}
{\cal C}_T=\int{\rm d}\Omega\frac{{\rm d}\sigma_T}{{\rm d}\Omega}\{G[1^\prime 
2^\prime]
-G[12]\}
\label{eq:colltur}
\end{equation}
that takes into account the correlations via the turbulent so far unspecified 
correlation functional $G[12]$. In 
this expression the positive definite turbulent differential cross-section 
$\sigma_T$ contains the supposed interactions between the clustered groups of 
particles at different scales formally in the same way as in 
Boltzmann's case. The explicit form of the cross-section is not of primary 
importance for our purposes in the following. Its calculation for particular 
cases will be delayed to elsewhere.
Its nature to be a collisional cross section guaranties its positive 
definiteness. This is all
we need here in developing an equilibrium theory. 

The turbulent conglomerates behave like superparticles. If this is the case, one 
can model 
the two-particle correlation 
function as the uncorrelated product of functionals $g[f_1]$ of the one-particle 
distribution function tentatively describing the approximately uncorrelated 
interaction of the superparticles formed by the long range correlation of the 
particles in the system. The underlying assumption in this approach is that the 
long-range correlations become very weak on a sufficiently large scale such that 
the molecular chaos assumption holds again if the scale is chosen sufficiently 
large. In other words, the assumed scale-invariance must be broken at 
sufficiently large scales. This is certainly a valid assumption for most 
physical systems. Then we can write
\begin{equation}
G[12]\simeq g[f_1(1^\prime)] g[f_1(2^\prime)] -g[f_1(1)] g[f_1(2)].
\label{eq:gfunc}
\end{equation}
In accord with the previous discussion, higher order 
correlations are neglected in this representation. 
The so far unspecified mechanism of how the cluster 
conglomerates and congregates
are formed and which part of the
phase space is occupied by each of them is thereby 
buried in the particular functional form of $g[f_1]$.

\section{Correlation Functional and Equilibrium Distribution}
Our primary task is to find an analytic expression for the functional $g[f_1]$. 
Clearly,
any rigorous theory should derive $g[f_1]$ from the Liouville equation or at 
least from the kinetic equation for the two-particle correlation function $f_2$, 
i.e. from the second-lowest order equation in the BBGKY hierarchy. Here we 
follow a more 
heuristic elementary approach. 

Let us assume homogeneous force-free equilibrium conditions such that ${\rm 
d}f_1/{\rm d} t=0$. The Boltzmann equation then requires that the collision term 
(\ref{eq:colltur}) vanishes identically. A sufficient condition for this to 
happen is that the correlation functional
\begin{equation}
G[12]=0. 
\end{equation}
This requirement does not depend on the explicit form of the differential cross 
section as long as the latter is positive definite, as we have assumed. 
It is thus a very 
general assumption holding for any equilibrium theory. 
Taking the logarithm of $G[12]$ yields
\begin{equation}
\ln\,g[f_1(1^\prime)]+\ln\,g[f_1(2^\prime)] =\ln\,g[f_1(1)]+\ln\,g[f_1(2)].
\label{eq:lng}
\end{equation}
This expression corresponds to Boltzmann's equilibrium condition with the only 
difference 
that it is 
now demanded to be satisfied by the functionals $g[f_1]$ instead of the 
distribution function $f_1$ itself. However, 
because the underlying interaction between the particles remains to be 
deterministic,  $g$ must be a 
function of the constants 
of motion only, viz. particle number and energy. 
Thus the general solution to the above equation is
\begin{equation}
\ln\,g[f_1({\bf p};\kappa)]= -\beta[({\bf p-p}_0)^2/2m -\mu],
\label{eq:pp}
\end{equation} 
where $\beta, \mu$ are two arbitrary real constants, and ${\bf p}_0$ is the 
initial 
average momentum of the particles of mass $m$ that they may have possessed in 
common. The constant $\mu$ somehow takes care of the conservation of the 
particle number. In addition we introduced the free parameter $\kappa$ 
on that $g[f_1]$ may also depend, $\kappa$ playing the role of a control 
parameter that characterises the particular nature of the correlations caused by 
the specific long range processes in the medium.
The functional $g$ hereby turns out to be an exponential function of the 
particle energy. In order to find its functional dependence on $f_1$, 
it should be 
inverted for $f_1$. This is not possible without any further assumptions. One of 
these assumptions can be based on our additional knowledge 
that in the limit of molecular chaos the Boltzmannian statistical mechanics 
must be recovered. We, hence, liberately demand that in the limit $\kappa\to 
\infty$ the functional $g[f_1]$ should become the identity functional 
\begin{equation}
\lim\limits_{\kappa\to\infty} g[f_1({\bf p};\kappa)]=\textsf{\textbf I}[f_1({\bf 
p};1)]=f_B({\bf p}).
\label{eq:boltzlim}
\end{equation}
This requirement can be taken as a guideline in the search for the functional 
form of $g[f_1]$. 

Let us assume that the one-particle distribution function has 
been normalized to one. This assumption is easy to satisfy for any given 
distribution function since the physical interpretation of the distributions is 
to be a probability distribution function. The simplest possible function 
$g[f_1]$ that satisfies the above condition is
\begin{equation}
g[f]=\exp\,\{\kappa[1-f^{-1/\kappa}]\}, \qquad \kappa\leq \infty,
\label{eq:gf}
\end{equation}
where we dropped the index 1 for simplicity, since from now on we will 
understand $f=f_1$ as the one-particle distribution function. It is easy to 
check that (\ref{eq:gf}) actually becomes the identity functional for the free 
parameter $\kappa\to\infty$. There may be other functional forms as well which 
satisfy (\ref{eq:boltzlim}). However, in view of the free parameter $\kappa$, 
Eq. (\ref{eq:gf}) is the simplest choice one can make. 

In order to find the equilibrium distribution function $f({\bf p},\kappa)$, we 
invert Equation (\ref{eq:gf}) and combine the result with Equation 
(\ref{eq:pp}). This procedure yields
\begin{equation}
f({\bf p},\kappa)=A_\kappa\left\{1+\frac{\beta}{\kappa}[({\bf 
p-p}_0)^2/2m-\mu]\right\}^{-\kappa},
\label{eq:kappadistr}
\end{equation}
where $A_\kappa$ takes care of the proper normalisation of the distribution 
function. Taking the limit $\kappa\to\infty$ one again easily checks that 
$\lim_{\kappa\to\infty} f\to f_B$, and this 
distribution becomes the Boltzmann function in the large-$\kappa$ limit. Hence, 
$f({\bf p},\kappa)$ from (\ref{eq:kappadistr}) solves the equilibrium 
one-particle kinetic equation (\ref{eq:liou}) with the new collision term 
(\ref{eq:colltur}) including turbulent correlations. 

The new distribution 
function is a $\kappa$-generalised Lorentzian function similar to those particle 
distributions that have been measured in space plasmas \cite{cri88,cri91}. It may 
also apply to L\'evy flight statistics. It contains three free parameters, 
$\kappa, 
\beta$ and $\mu$. The limit $\kappa\to\infty$ suggests that $\beta=1/k_BT$ can 
be identified with the kinetic temperature of the system under consideration, 
while the so far unspecified free parameter $\mu$ is the chemical potential. For 
$\kappa\ll\infty$ this chemical potential itself becomes a function of $\kappa$. 
Moreover, for finite $\kappa$, unlike the Boltzmann case, it does not drop out 
of the distribution function but is retained suggesting that finite-$\kappa$ 
systems do not behave like normal statistical systems but possess an internal 
chemical potential that affects the addition or extraction of particles.

Before proceeding to the next sections where we prove that the above 
distribution 
function $f({\bf p},\kappa)$ describes a real thermodynamic equilibrium, we note 
that the free parameter $\kappa$  necessarily satisfies a number of 
restrictions imposed on it by the requirement that the equilibrium distribution 
function $f({\bf p},\kappa)$ must conserve the particle number and average 
energy of the particles in the system. A naive inspection of 
(\ref{eq:kappadistr}) may suggests that $\kappa >1$. Below we will show that 
actually $\kappa >5/2$ for an ideal gas, which is a much stronger restriction on 
$\kappa$. We also note once more that $\kappa$ implicitly contains all 
information about the real turbulent dynamics of the system. Calculation of 
$\kappa$ from elementary processes thus is a very difficult task. Some effort 
has been given to it in a simple plasma interaction model \cite{has85} where it 
had been shown that $\kappa$ was a function of the nonlinear dispersion relation 
of the plasma. However, a perturbation approach as used there may be justified 
only in very particular cases.

\section{The Meaning of the Correlation Functional}
In this section we investigate the nature of the somewhat mysterious correlation 
functional $g[f]$. 
Without the construction of either the full microscopic turbulent interaction 
theory or solution of the 
next higher order kinetic equation in the BBGKY hierarchy it is hardly possible 
to precisely determine 
why $g[f]$ has just the structure as given in Eq. (\ref{eq:gf}). We can, 
however, show that $g[f]$ 
necessarily contains many correlations. We write
\begin{equation}
\ln g[f]=\kappa (1-f^{-1/\kappa}) =-\kappa\frac{1-f^{1/\kappa}}{f^{1/\kappa}}.
\end{equation}
Since $f$ is a probability distribution and $0<f<1$, the second form of this 
expression suggests the use 
of a simple summation formula in order to obtain
\begin{equation}
\ln g[f] = -\kappa\left[f^{1/\kappa}\sum\limits^\infty_{j=0} 
f^{j/\kappa}\right]^{-1}.
\label{eq:lngf}
\end{equation}
From here the correlation functional can be written as an infinite product
\begin{equation}
g[f] = \left\{\prod\limits_{j=0}^\infty 
\exp\left[f^{(j+1)/\kappa}\right]\right\}^{-\kappa}.
\end{equation}
This expression shows very clearly that $g[f]$ for finite $\kappa$ is a product 
of infinitely many 
exponentials of all powers of the distribution function. It therefore contains 
all kinds of correlations on 
all scales. On the other hand, returning to $\ln g[f]$, we may realise that Eq. 
(\ref{eq:lngf}) can be 
expressed as an infinite chain fraction. Because $f<1$, one can expand the chain 
fraction and rewrite it 
as
\begin{equation}
\ln g[f] \simeq 
-\frac{\kappa}{f^{1/\kappa}}\left[1-\sum\limits^\infty_{j=0}f^{(j+1)/\kappa}
\right],
\end{equation}
an expression that immediately yields
\begin{equation}
g[f]\simeq \prod\limits^\infty_{j=0}\left[\exp\left(\kappa 
f^{j/\kappa}\right)/\exp\left(\kappa f^{-
1/\kappa}\right)\right].
\end{equation}
In an expansion with respect to $f$ the leading term in this last expression is 
when retained only
\begin{equation}
g[f]\sim \exp (-\kappa/f^{1/\kappa}) \sim f^{1/\kappa}/\kappa.
\end{equation}
Using this expression in the collision term Eq. (\ref{eq:colltur}) we have
\begin{equation}
G[12]\sim \kappa^{-2}\left\{(f[1^\prime]f[2^\prime])^{1/\kappa} - 
(f[1]f[2])^{1/\kappa}\right].
\end{equation}
This approximate form shows that in the very lowest approximation the collision 
integral assumes a 
form similar to Boltzmann's taken to the power $1/\kappa$. Hence, even to the 
lowest approximation the 
correlation scale is modified in a $\kappa$-system. Of course the discussion 
presented here is only 
qualitative, and the last expression is not valid in the limit $\kappa\to\infty$ 
in which case one has to 
return to the full expressions given above. But it shows that even an expansion 
of the exponentials does 
not remove the difficulties introduced by the multi-scale correlations in the 
system. Hence, a 
perturbation analysis cannot reproduce neither the new 
distribution function nor any $\kappa$-distribution. A 
more fundamental microscopic theory must either take advantage of 
renormalisation group analysis 
methods or use finite particle number numerical simulations in order to 
elucidate the underlying 
physical interactions. These are obviously highly nonlinear and contain many 
interacting scales.

\section{H-theorem and Entropy}
Having derived the equilibrium distribution function, we proceed to demonstrate 
that the 
new correlation statistical mechanics also obeys a $H$-theorem. In analogy to 
Boltzmann we define the 
$H$-function as
\begin{equation}
H(t)\equiv \int {\rm d}{\cal V}\,f\ln\,g[f]=\int {\rm d}{\cal 
V}\,\kappa(f-f^{1-1/\kappa}),
\label{eq:h}
\end{equation}
with ${\rm d}{\cal V}$ the phase-space volume element. Differentiating with 
respect to time $t$, we arrive at
\begin{equation}
\frac{{\rm d}H}{{\rm d}t}=\int{\rm d}{\cal V}\,\left(\ln\,g[f]+
\frac{{\rm d}\ln\,g[f]}{{\rm d}f} 
\right)\frac{{\rm d }
f}{{\rm d} t}.
\label{eq:ht1}
\end{equation}
In equilibrium ${\rm d} f/{\rm d} t=0$, and hence ${\rm d}H/{\rm d} t=0, H=$ 
const assumes a constant value. It can also be shown that under time-dependent 
condition not deviating too far from equilibrium this value is monotonically 
approached, and ${\rm d} H(t)/{\rm d} t\leq 0$. 

Actually, inserting into (\ref{eq:ht1}) for the collision term from 
(\ref{eq:colltur}), making use of the symmetries and rearranging one finds after 
some amount of algebra
\begin{eqnarray}
&&\frac{1}{2}\frac{{\rm d}(H+H^\prime)}{{\rm d}t}=
\frac{1}{4}\int{\rm d}{\cal V}\,{\rm d}\Omega\frac{{\rm d}\sigma_T}{{\rm 
d}\Omega}( 
g[1^\prime] g[2^\prime]-g[1]g[2])\cr
&&\cr
&&\times\left\{ 1+ \left(\sum\limits_{i=1,2}-
\sum\limits_{i=1^\prime,2^\prime}\right)\frac{\delta}{\delta f(i)}\right\} 
\ln\left(\frac{g[1]g[2]}{g[1^\prime] 
g[2^\prime]}\right),
\label{eq:hfin}
\end{eqnarray} 
where $H^\prime$ is obtained from $H$ by exchanging primed and unprimed 
variables. The 
functional derivatives in (\ref{eq:hfin}) are simply equal to the powers 
$f^{-(1+1/\kappa)}$ of the 
corresponding distribution functions. It is then simple matter to show that the 
integrand in 
(\ref{eq:hfin}) can never be positive, and that $H$ is a monotonically 
decreasing function of time, if only 
$\kappa\leq \infty$, which proves the $H$-theorem for these values of $\kappa$. 
It is, however, important to note that the functional derivative term in 
(\ref{eq:hfin}) increases the negative derivative of the $H$ function 
and thus accelerates the tendency towards thermal equilibrium. This behaviour is 
not unexpected, 
because the long range correlations should enhance the dissipation of free 
energy. In other words, the 
correlations contribute an additional amount to the entropy. The system always 
tends towards equilibrium and never departs from it by itself unless it is 
disturbed from the outside and free energy is added to it. 

The $H$-theorem permits for an easy definition of the entropy ${\cal S}_T$ of 
the system. 
But before reading it from Eq. (\ref{eq:h}) we present some physical argument.
 
There are several ways of constructing the entropy of the system. For brevity 
we refer to Eq.\ (\ref{eq:pp}). Note that this equation is 
linear in the particle energy $\epsilon({\bf p})$ 
(for simplicity, we drop the translational momentum ${\bf p}_0$ at this place). 
Performing the ensemble average on both sides, we obtain
\begin{equation}
\langle\ln\, g[f]\rangle = -\beta\left(\langle\epsilon({\bf 
p})\rangle-\mu\right)= \ln\, 
g[f(\langle\epsilon\rangle)]
\label{eq:av}
\end{equation}
This result suggests that in analogy to the definition of Boltzmann's entropy 
(e.g., \cite{lan72}) the new turbulent
$\kappa$-entropy is defined as the ensemble average over the logarithm of the 
functional $g[f]$:
\begin{equation}
{\cal S}_T[f({\bf p};\kappa)]=-k_B\int\!{\rm d}^3{\bf p}f({\bf p};\kappa)\ln\, 
g[f({\bf 
p};\kappa)].
\label{eq:skappa}
\end{equation}
It thus turns out that the functional $g[f]$ plays the role of the probability 
distribution of states in the phase space that has  been deformed to some extent
by the action of the long-range correlations. This entropy is still an additive 
quantity, but it is $g[f]$ and not $f$ itself that determines the distribution 
of states. Inserting for $g[f]$, one immediately obtains the following 
representation of the entropy (actually, this is the entropy density of the 
system)
\begin{equation}
{\cal S}_T[f({\bf p};\kappa)]=-k_B\kappa\int\!{\rm d}^3{\bf p} f({\bf 
p};\kappa)(1- f({\bf 
p};\kappa)^{-1/\kappa}).
\label{eq:stsallis}
\end{equation}
Up to the dimensional factor $k_B$, this is just the negative of the 
$H$-function, and we conclude immediately 
that ${\rm d}{\cal S}_T/{\rm d}t \geq 0$. The entropy can only increase. 

In order to be in concordance with the second law, the entropy must in addition 
be a concave function. It is indeed simple matter to show that this is the case 
for the above definition of the entropy. Assume two independent distribution 
functions $f_1, f_2$ (indices do not refer to one- and two-particle 
distributions, here, but merely designate two different one-particle 
distributions). It then follows from Eq.\ (\ref{eq:skappa}) that
\begin{eqnarray}
{\cal S}_T\left[\frac{1}{2}(f_1+f_2)\right]&=&\frac{1}{2}\left({\cal S}_T[f_1]+
{\cal S}_T[f_2]\right)
+\frac{k_B\kappa}{2}\int{\rm d}^3{\bf p}\hfill\cr
&&\cr
&\times&\left[2^{1/\kappa}
(f_1+f_2)^{1-1/\kappa}-f_1^{1-1/\kappa}-f_2^{1-1/\kappa}
\right].
\end{eqnarray}
Remember that $0<f<1$.
The extra term on the right-hand side of this expression is always
positive which proves the concavity of the entropy.

The above definition of the entropy resembles Tsallis' \cite{tsa88} proposal of 
a non-extensive entropy. In fact, one can show that the two are indeed related. 
Assume $N_k$ 
particles to be distributed on $\tau_k$ quantum states in the phase space volume 
element, $\Delta 
p_k\Delta q_k$. The number of particles in state $k$ is $N_k=f(\kappa)\tau_k$. 
Resolving the integral in 
(\ref{eq:stsallis}) into a sum using this re-interpretation of $f(\kappa)$, the 
entropy becomes
\begin{equation}
{\cal S}_T(\kappa)=-k_B\kappa\sum\limits_k\tau_k\frac{N_k}{\tau_k}\left[1- 
\left(\frac{N_k}{\tau_k}\right)^{-1/\kappa}\right],
\label{eq:tsallis}
\end{equation}
which is a slightly rewritten version of Tsallis' definition of ${\cal S}_T$ 
for $\kappa<\infty$. In this representation Tsallis' distinction between his 
cases $q<1$ and $q>1$ disappears and leaves the only physically reasonable case 
$\kappa<\infty$. Moreover, because the entropy is the logarithm of the 
statistical weight, $\Omega_\kappa$, we can identify
\begin{equation}
\Omega_\kappa=\exp\left\{ -k_B\kappa\sum\limits_k N_k
\left[1-\left( \frac{N_k}{\tau_k}\right)^{-1/\kappa}\right]\right\}.
\label{eq:omega}
\end{equation}
as the fundamental `thermodynamic weight' function. As in conventional statistical 
mechanics, $\Omega_\kappa$ can be written in terms of a 
product of the statistical weights of the $k$ states
\begin{equation}
\Omega_{\kappa 
k}=\left\{\exp-\kappa\left[1-\left(\frac{N_k}{\tau_k}\right)^{-1/\kappa}
\right]\right\}^{N_k}.
\label{omegak}
\end{equation}
One easily confirms that for $\kappa\to \infty$ this becomes the correct 
Boltzmann statistical 
weight when identifying $\ln\,N_k^{N_k}$ as the large-$N_k$ limit of 
$\ln\,N_k!$. Hence, Eq.\ 
(\ref{omegak}) contains a large $N_k$ approximation. At the current stage it is 
not easily possible to resolve this approximation and to recover the correct 
form of the $\kappa$-statistical weight. The form of $\Omega_{\kappa k}$ given 
in (\ref{omegak}) shows that the state counting process underlying the 
$\kappa$-generalised statistical mechanics is highly complicated by the 
dominance of the correlations. Obviously, the correlations contract the phase 
space to a reduced volume that may consist in a strange attractor, for 
instance. The 
degree and process of contraction is hidden in the exact value of $\kappa$, as 
well as in the above statistical weight function.

Because $\Omega_{\kappa k}$ contains the information about the ways of how the 
particles can be 
distributed in the $k$ states (with degeneracy $\tau_k$), it also tells that for 
$\kappa\neq\infty$ the distribution procedure is no simple stochastic process. 
Interestingly, Eq.\ (\ref{eq:tsallis}) corresponds to
the definition of fractal dimensions through the natural measure of strange 
attractors $\nu_i=\lim_{t\to\infty}(\theta_i/t)$, where $\theta_i$ is the 
fraction of time
a certain volume element $i$ on the basin of the attractor is visited by the
orbits. Given a resolution $\eta$, one may write for the
multi-fractal dimension 
\begin{equation}
D_\kappa=\lim\limits_{\eta\to 0} \left[I(\kappa,\eta)/\ln(1/\eta)
\right],
\label{eq:dim}
\end{equation}
where we defined the information (neg-entropy) 
$I(\kappa,\eta)=-\sum_i^{N(\eta)}\nu_i\ln g[\nu_i]$,
and $N(\eta)$ is the number of intervals of lengths $\eta$ to cover
the attractor (cf., e.g., \cite{ott93,bad87}). Note that $\kappa$ is not a natural 
number here. Since $\nu_i$ is a probability measure,
one recognises that $N_k/\tau_k$ corresponds to the natural measure of the 
attractor, which in the continuous case is taken over by $f(\kappa)$. 
This points on a close connection between the underlying nonlinear dynamics 
and the Lorentzian statistical mechanics developed in the present
communication. For $\kappa\to\infty$ one recovers the 
information dimension 
\begin{equation}
D_{inf}\equiv D_\infty=\lim\limits_{\eta\to 0}\sum\limits_{i=1}^{N(\eta)}\nu_i\ln\nu_i
/\ln\eta,
\end{equation}
while due to the boundedness of $\kappa\geq\kappa_m$ from below,
the limit $\kappa\to 0$ does not exist. This implies that the box-counting
dimension is never realised. The multi-fractal dimensions are limited
from above by $D_\kappa \leq D_{\kappa_m}$.

\section{Ideal Gas}
With the above instrumentation at hand one can investigate the behaviour
of the ideal $\kappa$-gas. 
Normalising the distribution function Eq.\ (1) 
to the particle number density $N/V$, we find $A_\kappa={\cal Z}_\kappa^{-1}$ 
where 
\begin{equation}
{\cal Z}_\kappa =\frac{V}{N}\left(\frac{2\pi m\kappa}{\beta h^2}\right)^{3/2}
\frac{\Gamma(\kappa-3/2)}{\Gamma(\kappa)}\left(1-\frac{\beta\mu}{\kappa}
\right)^{3/2 -\kappa}.
\label{eq:partfunc}
\end{equation}
The latter function is interpreted as the grand partition function. It depends
on the particle number $N$, volume $V$, the chemical potential $\mu$ and 
temperature functional $\beta(T)$, $\kappa$ and on Planck's constant $h$. 
In order to find the entropy of the ideal gas one carries out the integration
in Eq. (\ref{eq:stsallis}). The result is
\begin{equation}
{\cal S}_T =(k_B\kappa)\left\{\frac{\kappa-1}{\kappa-5/2}
\left[\frac{\Gamma(\kappa-3/2)}{\Gamma(\kappa)}\left(1-\frac{\beta\mu}{\kappa}
\right)^{3/2 -\kappa}\frac{n_{Q\kappa}}{\langle n\rangle}\right]^{1/\kappa}
-1\right\},
\label{eq:sideal}
\end{equation}
where $\langle n\rangle =N/V$ is the known average particle density in the
volume, and 
\begin{equation}
n_{Q\kappa} = \left(\frac{m\kappa}{2\pi\beta\hbar^2}\right)^{3/2}
\end{equation}
is the $\kappa$-modified quantum density. It requires 
some nontrivial algebra to demonstrate that this expression actually
becomes the classical Boltzmann entropy in the limit $\kappa\to\infty$, and
that $n_{Q\kappa}$ tends to approach $n_Q$. The above expression also
suggests that in an ideal $\kappa$-gas the Boltzmann constant is modified
and depends on the correlation properties of the gas through $\kappa$.
This makes $\kappa$ a sensible function of the macroscopic properties
of the gas. In particular, $\kappa (T)$ will depend on the temperature 
$T$ because
it is suspected that for $T\to\infty$ the medium should become totally
disordered. In this case $\lim_{T\to\infty}\kappa\to\infty$, and classical
Boltzmann statistics will apply again.

In order to complete the formalism we identify the thermodynamic potential
of the ideal gas with
\begin{equation}
\Omega(\mu,T,V)=-\frac{V\langle n\rangle\kappa}{\beta}{\cal Z}_\kappa
=-k_B\kappa VT\langle n\rangle.
\label{eq:gibbs}
\end{equation}
It allow to calculate the average particle density from
\begin{equation}
\langle n\rangle =-\frac{1}{V}\left(\frac{\partial\Omega}{\partial\mu}
\right)_{TV}= n_{Q\kappa}\frac{\Gamma(\kappa -1/2)}{\Gamma(\kappa)}
\left(1-\frac{\mu}{\kappa k_BT}\right)^{5/2-\kappa}.
\end{equation}
However, the average density is a known quantity. Hence this is an 
equation for the chemical potential of the ideal gas. Inverting it for
$\mu$, we obtain finally
\begin{equation}
\mu=\kappa k_BT\left\{1-\left[\frac{n_{Q\kappa}}{\langle n\rangle}
\frac{\Gamma(\kappa -1/2)}{\Gamma(\kappa)}\right]^{2/(2\kappa-5)}\right\}.
\label{eq:chempot}
\end{equation}
From here all remaining thermodynamic quantities can be easily calculated.
The last expression is of two-fold interest. It shows that the chemical
potential of an ideal $\kappa$-gas is negative because the quantum
density is always considerably higher than the average density. Implicitly
we had made this assumption already in calculating the entropy when
we assumed that the phase-space integrals would converge when integrating 
over the interval $0\leq p<\infty$. Positive chemical potentials may not be 
excluded, however, in non-ideal gases. In such cases the theory becomes
naturally more involved.
The most important observation is that the control parameter $\kappa$
is bound from below by
\begin{equation}
\kappa > 5/2.
\end{equation}
This limitation requires that the flattest physically possible distribution functions
can only have slopes larger than 2.5 in energy space. It is interesting to 
note that observed distribution functions seem to satisfy this condition.
Extension to non-ideal gases are straightforward. One simply replaces
$\epsilon_{\bf p}={\cal H}$ with the full one-particle Hamiltonian ${\cal H}$
of the system including all interaction potentials. However, due to
the complicate form of ${\cal H}$, approximation methods are needed
in this case in order to calculate the thermodynamic functions and
velocity moments. Generalisation to the relativistic case requires
the substitution $\epsilon_{\bf p}=m\gamma_r({\bf p})c^2$, where
$\gamma_r$ is the relativistic factor. In this way the present theory
has all the potentialities of a kinetic theory. A further generalisation
to the quantum case \cite{tre98} will be published elsewhere.

\section{Conclusions}
We have presented a generalisation of the Boltzmann equation to some
special class of collision term that is not based on binary collisions. 
Rather it takes into account long-range correlations between the particles. 
The equilibrium state of such systems
is described by a generalised $\kappa$ Lorentzian distribution function 
with $\kappa$ not necessarily being a rational function and being limited from
below. It is also described by a particular form of the (turbulent)
entropy ${\cal S}_T$. We believe that in collisionless
systems close to criticality when long-range correlations dominate such
thermodynamic equilibria will be realised. In future work is
has to be shown, however, if and how the proposed collision term can
be derived from the BBGKY hierarchy. So far justification
\cite{cri88,cri91,lin95} is given mainly by the surprisingly 
frequent experimental observation of $\kappa$-like distributions in
collisionless space plasmas. Their high-energy tail slope
generally agrees with the theoretical bound on $\kappa$ found here.
Observed deviations at lower energies may be the result of
different effects, not the least is the appearance of the chemical
potential $\mu$ in the theoretically correct distribution which is
not taken care of in the experimental fits.

The present analysis leaves open a large number of questions that
are related to the formal application of the Lorentzian statistical mechanics
to non-equilibrium processes, to its microscopic justification, as well as to
the identification of the underlying process of state countings. At the
present state of the theory it seems not easily possible to conclude
about the counting procedure. Clearly, the correlations acting on all scales
do not permit for a simple random distribution of states as in Gibbs-Boltzmann
statistical mechanics. It may be suspected that therefore the phase-space will not
homogeneously be visited by the system under such conditions. Only the strange attractor
is visited frequently during the evolution of the system while large holes
may exist in phase-space which are never even touched. In quasi-equilibirum the system
will repeat to circulate on the strange attractor but even its coverage will not
be homogeneous during the available time $t<\tau_c$. It thus does 
not so much surprise that the statistical weight $\Omega_{k\kappa}$ 
turns out to be an extremely complicated expression. 

As has been pointed out earlier in this paper any microscopic justification
of the new statistical mechanics must identify the non-linear process. A more
formal possiblity is to return to the second-lowest kinetic equation in the
BBGKY hierarchy, the equation for the two-particle distribution function
$f_2[12]$, and to solve it under appropriate assumptions as for instance
the neglect of collisions and correlations in this state. Such a solution should
probably yield a non-trivial collision integral that in the most general case
will include correlations. It should be of considerable interest to determine,
under which conditions this collision term can be brought into the form
heuristically proposed in the present communication. Another possibility
is to use few-particle numerical simulations and to determine the structure of
the distribution function when performing very many experiments with
slightly different initial conditions.

One of the fundamental questions concerns the meaning of the turbulent
entropy ${\cal S}_T$. Actually, the letter version of the present paper
has been rejected from publication in Physical Review Letters with the
argument that it would make no sense to define a new entropy. It is
clear, of course, that there is only one entropy in the Universe, viz.
disorder. However, the mathematical description of disorder can nevertheless
require different expressions to describe entropy growth in different
phases of the evolution of the system. In particular, under conditions when
long-range correlations, scale invariance, or strong nonlinearity dominate
over collisionality entropy may obey different laws than in Boltzmann 
statistical mechanics. 

Aside from these basic questions the structure of the new kinetic theory
and the quasi-equilibrium distribution function lead to the important questions
of the corresponding non-equilibrium and transport theories. The boundedness of
the control parameter $\kappa$ from below implies that only a limited number of
fluid moments can in practice be determined from the distribution. What does this 
limitation imply? Does it hint on a self-closure of the turbulent system or
does it force us to invent a new kind of moment calculation? Intuitively it
seems to be clear that the particle number in the energetic tail of the
Lorentzian distribution must be limited in order to avoid a ultra-violet
catastrophe. But what does it mean in reality? Obviously the number of
particles in the tail is subject to self-limitation. Too many energetic particles
may drive the system into a stronger turbulence causing stronger scattering
of the particles which may lead to bulk heating and otherwise limitation of
the energetic particle flux. It is important to identify the actual process
that governs this limiting cycle. Only when this problem has been
solved, a reliable transport theory can be developed. One of its ingredients
is the turbulent cross section. We have shown that it is not important to
be known for the determination of the equilibrium properties of
the Lorentzian gas. Knowing the distribution function the average
cross section can be calculated subsequently. This will be done in
future work. 

We finally point on the many possiblities of applications of the new theory to
collisionless thermodynamics, equilibria and stability processes, turbulence, 
radiation, and its extension to the quantum regime. Some of these questions
will be investigated elsewhere.

\section*{Acknowledgments}
I thank A. Kull for many enlightening discussions, J. Scudder, G.
Morfill, J. Geiss, B. Hultqvist, 
C. McKee, F. Bertoldi, M. Scholer and R. Durisen for interest
as well as J. Geiss, B. Hultqvist and G. Morfill for moral support.
This work has been performed as part of a senior visiting scientist
program at the International Space Science Institute in Bern, after it had
been initiated during a visiting professorship 
at the Solar Terrestrial Environmental Laboratory of Nagoya University, Japan. 
The hospitality of Y. Kamide and S. Kokubun at STEL, Toyokawa, as 
well as the support of Nagoya University is gratefully acknowledged.

\end{document}